\documentclass[prl,twocolumn,floatfix,showpacs,amsmath,amssymb] {revtex4}

\usepackage{graphicx}

\begin{document}

\title{Signatures of Dynamical Heterogeneity in the 
Structure of Glassy Free-energy Minima}
\author{Pinaki Chaudhuri$^{1,2}$, Smarajit Karmakar$^{1}$ and Chandan Dasgupta$^{1}$}

\affiliation{$^1$ Centre for Condensed Matter Theory, 
Department of Physics, Indian Institute of Science, Bangalore 560012, India.\\
$^2$ LCVN, Universit\'e Montpellier II, 34095 Montpellier, France.} 

\begin{abstract}
From numerical minimization of a model free energy functional for a system of
hard spheres, we show that the width of the local peaks of the time-averaged
density field at a glassy free-energy minimum exhibits large spatial variation,
similar to that of the ``local Debye-Waller factor'' 
in simulations of dynamical
heterogeneity. Molecular dynamics simulations starting from a particle 
configuration generated from the density distribution at a glassy free-energy
minimum show similar spatial heterogeneity in the degree of localization,
implying a direct connection between dynamical heterogeneity and 
the structure of glassy free energy minima.
\end{abstract}


\maketitle

Observation of dynamical heterogeneity, both in experiments~\cite{expt} and 
in simulations~\cite{sim}, 
has been a significant step in the attempt to understand the behaviour of
glass forming liquids. The degree of 
spatial variation of the ``propensity for motion''~\cite{harrowell1} 
of individual
particles, defined as the mean-square displacement of a particle
from its initial position,
provides a direct measure of dynamical heterogeneity. 
The local Debye-Waller factor (short-time rms displacement from the average
position) of individual particles in the initial configuration has been found
\cite{harrowell} to be strongly correlated with the spatially heterogeneous 
propensity for motion over longer time scales.
However, subsequent work~\cite{appign} has 
shown that this correlation exists only
for time scales shorter than the $\alpha$-relaxation time.
Spatial heterogeneity of the local Debye-Waller factor has also been 
observed~\cite{volm} in simulations 
below the glass transition temperature.

The physical origin of dynamical heterogeneity is
not well-understood at present. 
In particular, it is not clear whether the occurrence of
spatially heterogeneous dynamics can be explained within 
the ``free-energy landscape'' description~\cite{wolynes,cdotv,parisi}
of glassy behavior in which the complex dynamics is attributed to the presence
of a large number of ``glassy'' local minima of the free energy.
Density functional theory (DFT)~\cite{ry}, in which the free energy
is expressed as a functional of the time-averaged local density, 
provides a convenient framework for exploring the free-energy landscape. 
In this
description, a glassy free-energy minimum is a local minimum 
of the free energy 
functional with a strongly inhomogeneous but non-periodic density distribution.
The local peaks of the density distribution represent 
the time-averaged positions
of the particles, and the width of a local peak is analogous to
the local Debye-Waller factor measured in simulations.
In this description, the $\alpha$-relaxation time
corresponds to the time scale of transitions between different glassy
minima~\cite{wolynes,cdotv}. 
Therefore, the density distribution at a typical glassy 
free-energy minimum should
correspond to an average of the local density over a 
time scale shorter than the
$\alpha$-relaxation time. If
this description is valid, then the spatial variation of the propensities for
motion observed in simulations over time scales shorter than the
$\alpha$-relaxation time (which, as discussed above, 
is strongly correlated with 
the spatial variation of the local Debye-Waller factor) 
should be manifested in the
structure of glassy free-energy minima as a similar spatial variation of the
widths of the local peaks of the density distribution. The observation of such
spatial variation of the peak width would, therefore, provide an explanation of
dynamical heterogeneity within the ``free-energy landscape'' description. This
would be an alternative to the recently proposed ``mode-coupling'' 
description~\cite{bb} of dynamical heterogeneity.

In this Letter, we have used numerical minimization of the 
Ramakrishnan-Yussouff (RY) free energy functional~\cite{ry} 
for a hard sphere system
to study the density distribution at glassy free-energy minima. Using a 
Gaussian superposition approximation~\cite{singh,das,kim}, 
as well as unconstrained numerical
minimization of a discretized version of the free-energy 
functional~\cite{cdotv},
we show that the width of the local density peaks 
at glassy minima exhibits large spatial variation
similar to the spatial heterogeneity of the local Debye-Waller factor found in
simulations~\cite{harrowell,volm}. We have also performed
molecular dynamics (MD) simulations
starting from a particle configuration corresponding to a glassy 
minimum. The simulation results for the rms displacement
of the particles from their average positions 
are found to be strongly correlated with the corresponding
widths of the local density peaks obtained in the DFT
calculation. These results establish a direct connection between 
dynamical heterogeneity and
the structure of glassy free-energy minima.
 
The RY free energy functional\cite{ry} for a system of hard spheres
has the form 
\begin{eqnarray}
\beta F &=& \int{d {\bf r}\{\rho({\bf r})
\ln (\rho({\bf r})/\rho_0)-\delta\rho({\bf r})\} }  \nonumber \\
&-&\frac{1}{2}\int{d {\bf r} \int {d{\bf r}^\prime
C({|\bf r}-{\bf r^\prime|}) \delta \rho ({\bf r}) \delta
\rho({\bf r}^\prime)}} ,
\label{ry0}
\end{eqnarray}
where $\beta=1/(k_B T)$, $T$ is the temperature, $\delta \rho ({\bf r})
\equiv \rho({\bf r})-\rho_0$ is the
deviation of the time-averaged local density ${\rho(\bf r})$ from 
the density $\rho_0$ of the uniform liquid, and $C(r)$ is the
direct pair correlation
function of a uniform hard-sphere 
liquid at density $\rho_0$. In Eq.(\ref{ry0}),
we have taken the zero of the free energy at its uniform liquid value.
In earlier studies~\cite{singh,das,kim} of glassy minima of this free energy, 
the local density $\rho({\bf r})$
was approximated as a superposition of Gaussians,
$\rho({\bf r})=(\frac{\alpha}{\pi})^{\frac{3}{2}}\sum_{i=1}^{N}
\exp[-\alpha({\bf r}-{\bf R}_i)^2]$, 
with the centers of the Gaussians, $\{\bf{R}_i\}$, forming a fixed
amorphous structure which was taken to be 
either a random close packing 
of hard spheres~\cite{singh,das}, or a
particle configuration from MD simulations~\cite{kim}. The free energy was then
minimized with respect to the single variational parameter $\alpha$.
In all these calculations, glassy states with 
free energy lower than that of the uniform liquid were 
found at high densities. 

\begin{figure}[htbp]
\includegraphics[scale=0.6,angle=-90]{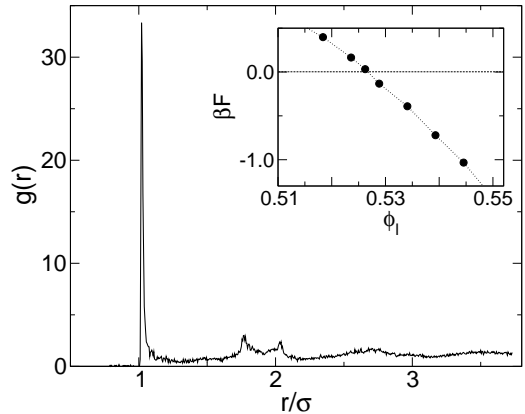} 
\caption{The pair distribution function $g(r)$ of a glassy minimum 
at $\phi_l=0.526$ obtained using the Gaussian superposition 
approximation and the HG form of $C(r)$. The inset shows the
variation of the dimensionless free energy (${\beta}F$) of the glassy minimum
with the reference liquid packing fraction $\phi_l$.} 
\label{fig1}
\end{figure}

Since we want to explore the possibility of 
the degree of localization (measured
by the width parameter $\alpha$) being different for different particles, we
have assumed the following form for the density profile:
\begin{equation}
\rho({\bf r})=\sum_{i=1}^{N}\frac{(1+\eta_i)\rho_0V}{N}
\left(\frac{\alpha_i}{\pi}\right)^{3/2}\exp[-\alpha_i({\bf r}-{\bf R}_i)^{2}]
\label{ry1}
\end{equation}
where $\alpha_i$ and $\eta_i$ are, respectively, parameters that
characterize the width and the height of the peak of the density profile 
at the point ${\bf R}_i$. In our 
calculations, initial values of $\{\bf{R}_i\}$ are taken from particle 
configurations obtained from MD 
simulations of $N=500$ hard spheres.  
We then minimize the RY functional with respect to 
the $2500$ parameters,
$\{\bf{R}_i\}$, $\{\alpha_i\}$, $\{\eta_i\}$, and also with respect to
the volume $V$ at a reference liquid packing fraction 
$\phi_l \equiv\pi \rho_0\sigma^3/6$, 
$\sigma$ being the
hard sphere diameter.  
We use the Hendersen-Grundke (HG) expression~\cite{hg} for 
$C(r)$, as has been done
in previous calculations~\cite{singh,das,kim}.
The minimization leads to structures similar to glassy states observed in
simulations and experiments.  
In Fig.\ref{fig1}, the pair distribution
function, $g(r)$, of the set of coordinates $\{\bf{R}_i\}$ 
for one such minimum
has been plotted -- the split second peak 
of the $g(r)$ clearly indicates that the
corresponding structure is amorphous~\cite{john}.
The inset of Fig.\ref{fig1} shows the dependence of the free energy 
of this minimum on the packing fraction $\phi_l$.
The free-energy of the glassy minimum becomes lower
than that of the uniform liquid (i.e. $\beta F$ becomes negative)
at  $\phi_l=0.526$,
which, as expected, is lower than the value obtained in Ref.\cite{das,kim}. 

\begin{figure}[htbp]
\includegraphics[scale=0.3,angle=270]{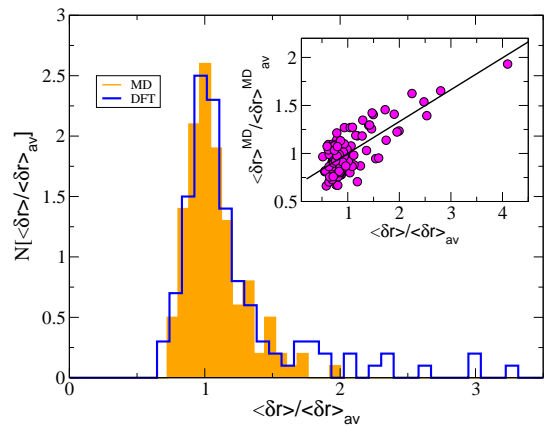}
\caption{Distribution of $\langle{\delta}r\rangle$, scaled by its
average value, for a glassy minimum 
at $\phi_l=0.55$ obtained
using the Gaussian approximation and the HG $C(r)$. 
The distribution of the rms displacement
$\langle \delta r \rangle^{\hbox{MD}}$ 
obtained from MD simulations is also shown for comparison. Inset: 
correlation between $\langle \delta r \rangle^{\hbox{MD}}$ and
$\langle \delta r \rangle$ (the straight line is a guide to the eye).}  
\label{fig2}
\end{figure}

In order to measure the widths of the local peaks of the density field at 
a glassy minimum, we have evaluated the
quantities $\langle{\delta}r\rangle_i=(\int_{v_i}d{\bf r}\,
|{\bf r}-{\bf R}_i|^2\rho({\bf r}))^{\frac{1}{2}}$, 
where $v_i$ is a small volume
centered at the point ${\bf R}_i$ (position of the $i$th local peak),  
such that $\int_{v_i}d{\bf r}\,\rho({\bf r})=1$.
We find that, indeed,  the values of $\langle{\delta}r\rangle_i$ are 
widely distributed, as can be seen from
Fig.\ref{fig2}, where we have plotted the distribution of 
$\langle{\delta}r\rangle$, scaled by its spatial average 
$\langle{\delta}r\rangle_{\hbox{av}}$, for a glassy minimum
at $\phi_l=0.55$. 
The distribution in Fig.~\ref{fig2} is qualitatively 
similar to those of the dynamical propensity and the local Debye-Waller factor
obtained in simulations~\cite{harrowell1,harrowell,volm}. 

To make a direct comparison between these DFT results and the dynamics 
of the system, we have performed MD simulations of the 
dynamics of the system near a free energy minimum. 
In these simulations, the values of $\{{\bf R}_i\}$ 
obtained in the  free-energy minimization are taken to be 
the initial particle 
positions. A few occurrences of neighboring local density peaks separated by 
less than $\sigma$ 
are removed using a local relaxation procedure~\cite{relax}.
Starting from the same initial configuration but using different sets of 
initial velocities randomly assigned 
from the Maxwell-Boltzmann distribution,
we have simulated the dynamics for $10^5$ collisions per particle, which 
corresponds to the middle of the plateau of the 
mean-square displacement of the particles from their initial positions.
Using $100$ such different trajectories, we 
have calculated the rms displacements,
$\{\langle \delta r \rangle_i^{\hbox{MD}}\}$, of all the particles from
their average positions. As shown in 
Fig.~\ref{fig2}, the distribution of $\langle \delta r \rangle^{\hbox{MD}}$
scaled by its average over all the particles is nearly identical to 
the distribution of $\langle \delta r \rangle$ obtained in the DFT calculation.

To test whether particles that have (small) large values of $\langle \delta r
\rangle$ in the DFT calculation show (small) large values of the rms
displacement in the MD simulation, we have performed free-energy minimizations
starting from 15 particle configurations obtained in the MD simulation.
To account for small differences between the initial particle positions
and the positions of the local density peaks at the corresponding free-energy 
minimum, we have divided the system into 125 ``cells'' of equal volume and
averaged both the DFT and MD values of $\langle \delta r \rangle$ over the 
particles lying in each cell. The inset of Fig.~\ref{fig2} illustrates the
correlation between these spatially ``coarse-grained'' values of 
$\langle \delta r \rangle$ obtained from DFT and MD calculations. The degree
of correlation is quite large, with a correlation coefficient
of about 0.76. The correlation coefficient is larger ($\simeq 0.85$) for
the subset consisting of
10\% of data points with the highest values of $\langle \delta r \rangle /
\langle \delta r \rangle_{\hbox{av}}$.
 
Next, we have removed the Gaussian constraint on the density field and 
obtained glassy free-energy minima using a
numerical scheme\cite{cdotv} for minimizing a 
discretized version of the RY functional for hard spheres. 
To discretize the RY functional,
a simple cubic computational mesh of size
$L^3$ and with periodic boundary conditions is introduced.
On the sites of this mesh, we define density variables
$\rho_i \equiv \rho({\bf r}_i) h^3$, where $\rho({\bf r}_i)$ is the
density at site $i$ and $h$ the spacing of the
computational mesh. 
The free energy of Eq.(\ref{ry0}) 
is then numerically minimized as a function of 
the density variables~\cite{cdotv}.

\begin{figure}[htbp]                                                           
\includegraphics[scale=0.6,angle=-90]{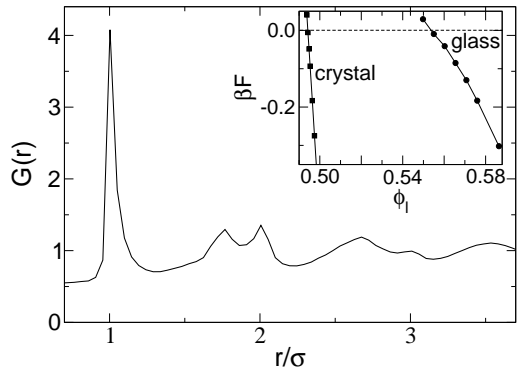}
\caption{The correlation function $G(r)$ of the time-averaged local density
for a glassy minimum at $\phi_l=0.586$, obtained from 
unconstrained minimization using the PY $C(r)$. 
The inset shows the variation of the free energy of the glassy minimum and that
of the fcc crystal with the packing fraction $\phi_l$.}
\label{fig3}
\end{figure}

In the numerical minimization, we use the Percus-Yevick (PY) expression 
for ${C(r)}$\cite{percus} because it 
leads to (see below) a very accurate value of
the crystallization density.
The input density field $\{\rho_i\}$ for the minimization 
is obtained using configurations from a MD simulation, for a
mesh-spacing $h \simeq 0.05{{\sigma}}$.
From the minimization of the free energy as a function of
$4.096\times10^6$ density variables, we have again
obtained local minima with glassy $\{\rho_i\}$.
The structure of a local minimum can be characterized by the two-point 
correlation function $G(r)$ of the {\it time-averaged} 
local density variables $\rho_i$ at the minimum (this function is different
from the pair distribution function $g(r)$ of Fig.~\ref{fig1}). 
In Fig.\ref{fig3}, 
we have plotted the $G(r)$ for a glassy free-energy
minimum at $\phi_l=0.586$. 
The glassy nature of the density distribution is indicated
by the split second peak of $G(r)$ and the positions of the two 
sub-peaks~\cite{john}.
The inset of Fig.\ref{fig3} shows 
the variation of the free energy
with the packing fraction $\phi_l$, which indicates that 
the free energy of the glassy 
minimum crosses that of the liquid
at $\phi_l=0.554$ which is very close to the packing fraction at the ideal
glass transition of mode-coupling theory. 
Also shown in the plot 
is the free energy of the fcc crystalline minimum, which becomes negative at
$\phi_l \simeq 0.497$, a value very close to the known 
packing fraction at crystallization.

\begin{figure}[htbp]
\includegraphics[scale=0.3,angle=-90]{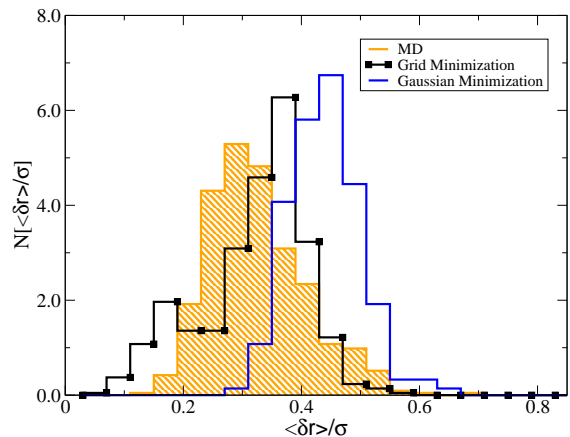} 
\caption{Distribution of $\langle{\delta}r\rangle$ 
at a glassy minimum
obtained using unconstrained minimization with the PY $C(r)$ 
for $\phi_l=0.613$. The 
distribution of the same quantity obtained using the Gaussian approximation 
with the same $C(r)$ 
and the distribution of $\langle \delta r \rangle^{\hbox{MD}}$ obtained
from MD simulations are also
shown for comparison.}
\label{fig4}
\end{figure}

Using the values of $\{\rho_i\}$ at a glassy minimum, 
we have identified 
the local density peaks and 
calculated their widths $\{\langle{\delta}r\rangle_i\}$, 
as in the calculation using the Gaussian approximation.  
It is clear from Fig.\ref{fig4}, which shows the distribution
of $\langle{\delta}r\rangle$ for a minimum at $\phi_l=0.613$, 
that glassy minima obtained using
this scheme also exhibit
considerable spatial heterogeneity in the values of $\langle{\delta}r
\rangle$. To relate these results to the dynamics, we have performed MD 
simulations, as described above, starting from the particle configuration
obtained from a free-energy minimum found in the unconstrained
minimization. These runs consist of 
50 different trajectories, each for
$10^5$ collisions per particle. 
As shown in Fig.~\ref{fig4},
the distribution of the rms displacements obtained from these MD runs is quite
similar to the distribution of $\langle \delta r \rangle$ obtained in the 
DFT calculation. 
The means and standard deviations of the two distributions, $0.35\sigma$ 
and $0.08 \sigma$, respectively, from MD and
$0.34\sigma$ and $0.09 \sigma$ from DFT, are nearly the same. 
Typical values of $\langle \delta r \rangle_i$
obtained using the HG $C(r)$ are about three times smaller than those
shown in Fig.~\ref{fig4}. Similar unrealistically small values of the
peak widths were obtained in earlier DFT calculations ~\cite{das,kim}
using the HG $C(r)$. We expect the
results obtained using the PY $C(r)$ to be more reliable. 

\begin{figure}[htbp]
\vspace{-0.2cm}
\includegraphics[scale=0.3,angle=-90]{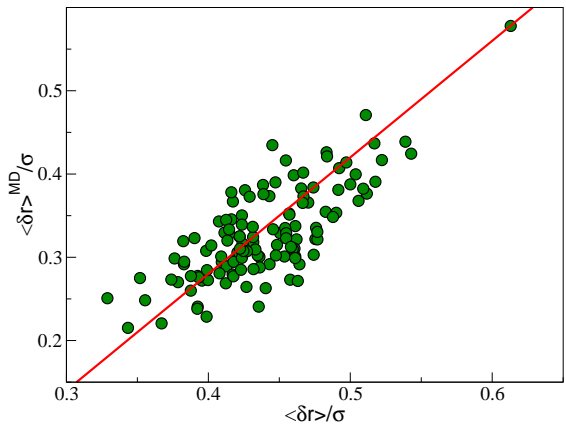} 
\caption{Correlation of the $\langle{\delta}r\rangle$ for a 
glassy minimum at $\phi_l=0.613$ (obtained using the Gaussian approximation
with the PY $C(r)$, see text) with the rms displacement 
$\langle{\delta}r\rangle^{\hbox{MD}}$ obtained from MD simulations. The 
straight line is a guide to the eye.}
\label{fig5}
\end{figure}

Since the local density peaks at the free-energy
minimum obtained using the PY $C(r)$ are much broader, 
the process of elimination of 
neighboring peaks separated by less than $\sigma$ causes substantial 
rearrangement of the particles. For this reason, a particle-by-particle 
comparison of these DFT results with the corresponding 
MD results is not possible. To make such a comparison, we have done a
new calculation using the Gaussian approximation, 
but this time with the PY $C(r)$,
to locate a free-energy minimum in which the local density peaks
coincide with the particle positions in the 
initial configuration used in the MD runs. In this
calculation, the density profile is assumed to have the form of Eq.(\ref{ry1})
and the free energy is minimized with respect to the parameters $\{\alpha_i\}$
and $\{\eta_i\}$, keeping $\{{\bf R}_i\}$ fixed.
Fig.~\ref{fig5} shows the correlation between the 
$\{\langle \delta r \rangle_i\}$ obtained from this DFT calculation and 
the $\{\langle \delta r \rangle_i^{\hbox{MD}}\}$ obtained
from MD simulations (both spatially ``coarse-grained'' using 125 cells
as before). This plot shows a high degree of correlation, with a 
correlation coefficient of 0.75 (it increases to 0.87 for 
10\% of data points with the highest
values of $\langle \delta r \rangle$).
Also, as shown in Fig.\ref{fig4}, the 
distribution of $\langle \delta r \rangle$ obtained from this Gaussian DFT 
calculation is similar to those obtained from the other calculations.
These results and those shown in Fig.~\ref{fig2} 
establish a direct connection between
dynamical heterogeneity and the structure of glassy free-energy minima by 
showing that if a small region of a free-energy minimum contains local 
density peaks with small (large) widths, then 
the particles in the same region are very likely to exhibit small (large)
values of rms displacement for dynamics near this minimum.

To summarize, using DFT, we have shown that at a glassy 
minimum of the free-energy 
of the hard-sphere system, the 
degree of localization of the local particle density exhibits large spatial
variation. Also, by
carrying out MD simulations starting from 
a particle configuration corresponding to 
the density distribution at a glassy minimum, we have established 
that the time-averaged
density distribution at the minimum 
is closely related to the dynamical heterogeneity exhibited by the system.

We are grateful to Robin Speedy for useful discussions
and Dhrubaditya Mitra for help in computation. 
PC would like to thank SERC (IISc) for 
computation facilities and JNCASR for financial support.

\end{document}